\documentclass[preprint,aps,prd,showpacs,twocolumn,footinbib,10pt]{revtex4}
\usepackage{epsfig,epsf,graphics}
\usepackage{amsfonts,amssymb,amsmath}

\newcommand{\VEV}[1]{\left\langle #1 \right\rangle}
\newcommand{\lag}{{\cal L}}
\newcommand{\nn}{\nonumber}
\newcommand{\diag}{{\rm diag}}

\newcommand{\GSM}{SU(3)_C\times SU(2)_L\times U(1)_Y}
\newcommand{\Z}[1]{{\mathbb Z}_#1}
\newcommand{\supP}[1]{^{(#1)}}
\newcommand{\sla}[1]{#1\!\!\!/}

\newcommand{\bequ}{\begin{equation}}
\newcommand{\eequ}{\end{equation}}
\newcommand{\beqn}{\begin{eqnarray}}
\newcommand{\eeqn}{\end{eqnarray}}
\newcommand{\Ls}{\left(}
\newcommand{\Rs}{\right)}
\newcommand{\Ll}{\left[}
\newcommand{\Rl}{\right]}

\begin{document}

\preprint{\small KYUSHU-HET-130}

\title{Grand Gauge-Higgs Unification}

\author{Kentaro Kojima$^{1}$}
      \email{kojima@rche.kyushu-u.ac.jp}
\author{Kazunori Takenaga$^{2}$}
      \email{takenaga@kumamoto-hsu.ac.jp}
\author{Toshifumi Yamashita$^{3}$}
      \email{yamasita@eken.phys.nagoya-u.ac.jp}
\affiliation{$\rm ^1$Center for Research and Advancement in Higher Education,\\
 Kyushu University, Fukuoka 819-0395, Japan} 
\affiliation{$\rm ^2$Faculty of Health Science, Kumamoto Health Science University, 
 Izumi-machi, Kumamoto 861-5598, Japan} 
\affiliation{$\rm ^3$Department of Physics, Nagoya University, Nagoya 464-8602, Japan}%

\begin{abstract}
We propose a novel way to break grand unified gauge symmetries via 
the Hosotani mechanism in models that can accommodate chiral fermions. 
Adjoint scalar fields are realized through the so-called diagonal embedding method 
 which is often used in the heterotic string theory. 
We calculate the one-loop effective potential of the adjoint scalar field 
 in a five dimensional 
 model compactified on an $S^1/\Z2$ orbifold, as an illustration. 
It turns out that the potential is basically the same as the one 
 in an $S^1$ model, and thus 
 the results in literatures, in addition to the chiral fermions, can be realized 
 easily.
\end{abstract}



\maketitle

%
\section{Introduction}
%
The Hosotani mechanism~\cite{Hosotani} is one of the most 
 interesting features possessed by models with topological extra dimensions, and 
 has been studied extensively in the literature~\cite{S1, hierarchy, S1Z2}. 
In this mechanism, zero-modes of the extra-dimensional components of the higher
dimensional gauge fields acquire non-vanishing vacuum expectation 
values (VEVs) to break the gauge symmetry. 
Since the extra dimensional components behave as 
adjoint scalar fields at low energy~\cite{S1,hierarchy}, the mechanism has 
mainly been applied to the grand unified theories (GUTs) in the last century.
The idea, however, encounters the difficulty that one cannot obtain the chiral
fermions, and thus it is not phenomenologically viable.

After the work in Ref.~\cite{Kawamura} on the orbifold symmetry breaking within the 
 field theoretical framework, 
which opens a way to realize scalar fields in the fundamental representation 
from the extra-dimensional components of the higher dimensional gauge fields, 
the Hosotani mechanism is often applied to the electroweak symmetry 
breaking. In this 
scenario, one can obtain the chiral fermions, and in addition, 
it gives a new approach to the hierarchy problem~\cite{hierarchy} 
without using supersymmetry. 
The scenario is called gauge-Higgs unification and has been studied from various 
points of view~\cite{S1Z2}. 

It is natural to ask whether or not one can apply the Hosotani mechanism to the GUTs
without contradicting with the existence of the chiral fermion.  
Our aim in this letter is to establish a 
way to realize this application in the case of the orbifold compactification.
This is very attractive because the gauge symmetry-breaking patterns are 
completely determined by the calculable dynamics independently of details 
of the unknown ultraviolet completion, thanks to the finiteness of the Higgs 
effective potential~\cite{hierarchy}.
This is in great contrast to the orbifold breaking in which
the symmetry-breaking pattern is freely chosen by boundary conditions (BCs). 
In addition, at the same time, one can incorporate the chiral fermions easily.

There is, however, a difficulty in this attempt. 
The massless adjoint scalar fields, which originate from 
the extra-dimensional components, tend to be 
projected out at low energy in models that realize the chiral fermions.
This is caused by the difference between the BCs of the four-dimensional (4D) 
 vector components and those of the extra-dimensional ones. 
It is actually possible that they have a common BC, when there are  
directions compactified on a certain manifold, such as a torus. 
The chiral fermions, however, disappear when such directions exist.

This difficulty is shared with the heterotic string theory~\cite{hetero}. 
Since this theory is expected to contain the standard model, it has been thoroughly 
studied and a way to overcome the difficulty was proposed, through 
the so-called diagonal embedding 
method~\cite{DiagonalEmbedding, heteroticGUTs, IKMMTTTY}.  
This method extracts the diagonal part of $n$-copies of a gauge 
symmetry $G$, $(G)^n$, by, for instance, a $\Z{n}$ orbifold 
action that permutes them. 
In view of the orbifold projection, 
the eigenvalues of the permutation are given 
as $e^{2\pi i k/n}$ $(k=0, 1, \cdots, n-1)$. 
In particular, the gauge factor with $k=0$ corresponds to the diagonal part 
and remain unbroken by the orbifold action. In other words, the remaining gauge 
symmetry in the 4D effective theory is embedded into the diagonal part. 
A point is that if the phases of other eigenstates cancel those of the 
extra-dimensional components of the gauge field which comes from the 
geometrical twist, adjoint 
scalars appear in the massless spectrum~\cite{IKMMTTTY}.

We find no reasons that forbid the application of the same method 
to our phenomenological setup and we examine this possibility. 
An advantage of working in a simple field theoretical setup is that it is much 
easier to calculate the quantum corrections that tell us the positions of vacua.
By this, it is possible to determine the gauge symmetry-breaking patterns 
dynamically.
In this letter, as an illustration, we work on $S^1/\Z2$ models since 
extension to models with a more complex orbifold such as $T^2/\Z3$ is 
straightforward.
%
\section{Diagonal embedding in $S^1/\Z2$ model}
%

Let us rephrase the diagonal embedding method in terms of the BCs. 
In the $S^1/\Z2$ orbifold model, we should fix the BCs with respect to the 
reflections of the 5th-dimensional coordinate, $y$, at the two fixed points, 
 $y_0=0$ and $y_\pi=\pi R$. We choose them so that two-copies of the gauge 
symmetry $G$ are exchanged~\footnote{A similar BC is 
adopted in Ref.~\cite{ExchangeBC} to improve the naturalness and 
realize a dark matter candidate, at one of the fixed points.
}.
For this purpose, the Lagrangian should be symmetric under the exchange. 
Namely, we consider models with a $G\times G\times \Z2$ symmetry. 
We formally distinguish the two gauge factors with indices, as $G_1$ and $G_2$. 
Their gauge fields and generators are named as 
$A\supP1_M$ and $A\supP2_M$, and $T_1^a$ and $T_2^a$, 
respectively, where $M=\mu(=0\mbox{-}3),5$ is a 5D Lorentz index.

To be more concrete, we choose the BCs as 
\bequ
 A_\mu^{(1)}(y_i-y) = A_\mu^{(2)} (y_i+y), 
\eequ
 which are rewritten as
\bequ
 A_\mu^{(\pm)}(y_i-y) = \pm A_\mu^{(\pm)} (y_i+y), 
\label{eq:BCAmupm}
\eequ
where we have defined $X^{(\pm)}=(X\supP1\pm X\supP2)/\sqrt2$ 
and used $y_i(i=0,\pi)$.
Since $A_5$ has the parity opposite to $A_\mu$ to keep the Lagrangian 
invariant, they lead to 
\bequ
A_5^{(\pm)}(y_i-y) = \mp A_5^{(\pm)} (y_i+y).
\label{eq:BCA5pm}
\eequ
The BCs \eqref{eq:BCAmupm} imply that the gauge 
symmetries $G_1\times G_2$ are broken down to their diagonal 
part $G_\diag$ whose gauge field is $A_\mu^{(+)}$, while 
those in \eqref{eq:BCA5pm} show that the zero-mode of the extra-dimensional 
component $A_5^{(-)}$ exists, which behaves as the adjoint representation of 
$G_\diag$ as we will see soon. In this way, a massless 
adjoint scalar is obtained in the orbifold model. 

The Wilson line in this case is given as 
\begin{widetext}
\begin{eqnarray}
 W=\exp\Ls i\int_{0}^{2\pi R} g\Ls {A_5\supP1}^aT_1^a 
+ {A_5\supP2}^aT_2^a\Rs dy\Rs 
=\exp\Ls i\int_{0}^{2\pi R} \frac{g}{\sqrt2}
\Ls {A_5\supP+}^a(T_1^a + T_2^a) + {A_5\supP-}^a(T_1^a - T_2^a)\Rs dy\Rs,
\end{eqnarray}
\end{widetext}
 where $g$ is the common gauge coupling constant.
Taking the commutation relations with the generators of the diagonal group,
\bequ
 \Ll T_1^a+T_2^a, T_1^b\pm T_2^b\Rl = if^{abc}\Ls T_1^c\pm T_2^c\Rs,
\label{eq:commutation}
\eequ
we see that both combinations $\Ls T_1\pm T_2\Rs$ behave as adjoint 
representation under the diagonal group. Note that the 
normalizations of $\Ls T_1\pm T_2\Rs$ have no factor of $1/\sqrt2$ to 
satisfy the commutation relation \eqref{eq:commutation}. 
This results in the suppressed gauge coupling of $G_\diag$ compared to 
the original one, $g_\diag=g/\sqrt2$.

Since only the combination $A_5\supP-$ has zero-modes, we set the 
VEVs of $A_5$ as 
\bequ
\VEV{\frac{g}{\sqrt2}{A_5\supP-}^a} = 
\VEV{g{A_5\supP1}^a} = -\VEV{g{A_5\supP2}^a} = \frac{\theta^a}{\pi R}. 
\label{eq:BGVEV}
\eequ
Then the VEV of the Wilson line becomes
\bequ
\VEV{W}=\exp\Ls i\theta^a 2\Ls T_1^a- T_2^a\Rs \Rs.
\eequ
\section{Fermions}

Because of the $\Z2$ symmetry which is required to impose the BCs 
that exchange the two gauge groups, when we introduce a 
fermion, $\Psi\supP1({\bf R_1},{\bf R_2})$ with 
${\bf R_1}\neq{\bf R_2}$ 
where ${\bf R_1}$ and ${\bf R_2}$ denote its representations 
of $G_1$ and $G_2$, respectively, its $\Z2$ partner 
$\Psi\supP2({\bf R_2},{\bf R_1})$ should be also introduced. 
These fermions behave as a reducible representation ${\bf R_1}\times{\bf R_2}$ 
under the residual diagonal gauge groups. To keep 
the Lagrangian invariant, their BCs are given as
\bequ
 \Psi\supP1(y_i-y) = \eta_\Psi\gamma^5\Psi\supP2(y_i+y),
\eequ
where $\eta_{\Psi}$ is a parameter being $\eta_{\Psi}^2=1$.
Namely, 
\bequ
 \Psi\supP\pm(y_i-y) = \pm\eta_\Psi\gamma^5\Psi\supP\pm(y_i+y).
\eequ
This implies that a vector-like pair of the fermion zero-modes in the reducible 
 representation ${\bf R_1}\times{\bf R_2}$ appears. 

In the case with ${\bf R_1}={\bf R_2}\,(={\bf R})$, a single fermion 
 $\psi^{\alpha\beta}$ is allowed, where $\alpha$ and $\beta$ are the indices 
 of the representation ${\bf R}$ on which the elements of $G_1$ and $G_2$, 
 respectively, act.
It behaves as a reducible representation of the diagonal group: 
\bequ
 {\bf R}\times{\bf R} = \Ls{\bf R}_{s1}+\cdots\Rs + \Ls{\bf R}_{a1}+\cdots\Rs, 
\eequ
where ``$s$" and ``$a$" denote symmetric and 
anti-symmetric products, respectively. 
We define $\psi^{\alpha\beta}_\pm=(\psi^{\alpha\beta}\pm\psi^{\beta\alpha})/2$
so that $\psi_+$ and $\psi_-$ consist 
of ${\bf R}_{si}$ and ${\bf R}_{ai}$, respectively.

Its BCs are
\beqn &&
 \psi^{\alpha\beta}(y_i-y) = \eta_\psi\gamma^5\psi^{\beta\alpha}(y_i+y),
\\ \mbox{or}\quad &&
 \psi^{\alpha\beta}_\pm(y_i-y) 
 = \pm\eta_\psi\gamma^5\psi^{\alpha\beta}_\pm(y_i+y). 
\qquad
\eeqn
With this, we see that chiral zero-modes appear in this case~\footnote{
In addition, as usual, chiral fermions can be put on each brane at $y=y_i$. 
}.

The interaction terms among the fermions and the 5D gauge fields are written as 
\begin{widetext}
\beqn
&& g
 \Ls\begin{matrix}\bar\Psi\supP1 & \bar\Psi\supP2\end{matrix}\Rs
 \Ls\begin{matrix}
     {\sla A\supP1}^a T_{\bf R_1}^a + {\sla A\supP2}^a T_{\bf R_2}^a & 0 \\
     0 & {\sla A\supP1}^a T_{\bf R_2}^a+ {\sla A\supP2}^a T_{\bf R_1}^a
    \end{matrix}
 \Rs
 \Ls\begin{matrix}\Psi\supP1 \\ \Psi\supP2\end{matrix}\Rs
\nn\\
&&=\frac{g}{\sqrt2}
 \Ls\begin{matrix}\bar\Psi\supP+ & \bar\Psi\supP-\end{matrix}\Rs
 \Ls\begin{matrix}
     {\sla A\supP+}^a (T_{\bf R_1}^a + T_{\bf R_2}^a) & 
     {\sla A\supP-}^a (T_{\bf R_1}^a - T_{\bf R_2}^a) \\
     {\sla A\supP-}^a (T_{\bf R_1}^a - T_{\bf R_2}^a) & 
     {\sla A\supP+}^a (T_{\bf R_1}^a + T_{\bf R_2}^a)
    \end{matrix}
 \Rs
 \Ls\begin{matrix}\Psi\supP+ \\ \Psi\supP- \end{matrix}\Rs 
\nn\\
&&\ni \bar\Psi\supP\pm\frac{\theta^a}{\pi R}(T_{\bf R_1}^a - T_{\bf R_2}^a) 
             i\gamma^5\Psi\supP\mp. 
\eeqn
\end{widetext}
Here, $T_{\bf r}$ is the generator on the representation ${\bf r}$ and 
 $\sla A=A_M \Gamma^M$ 
 with $\Gamma^M=(\gamma^\mu,i\gamma^5)$.
Note that the last two expressions also hold for $\psi$, with the replacements 
 $\Psi\supP\pm\to\psi_\pm$ and $T_{\bf R_1}$ ($T_{\bf R_2}$) by $T_{\bf R}$ that 
 act on the first (second) index of $\psi_\pm$. 
Thus, we concentrate on $\Psi$ in the remaining in this 
section, except for specially mentioned comments.

On the background \eqref{eq:BGVEV}, the mass terms for the $n$-th Kaluza-Klein (KK) 
 modes of $\Psi\supP\pm$, with $n$ being a non-negative integer, are  
\begin{widetext}
\bequ
 \lag_{KK}= \frac{-1}R 
 \Ls\begin{matrix}\bar\Psi\supP{+}_n & \bar\Psi\supP{-}_n\end{matrix}\Rs_{\chi'}
 \Ls\begin{matrix}
     -n & 
     -i\theta^a (T_{\bf R_1}^a - T_{\bf R_2}^a)/\pi \\
     -i\theta^a (T_{\bf R_1}^a - T_{\bf R_2}^a)/\pi & 
     n
    \end{matrix}
 \Rs\gamma_5
 \Ls\begin{matrix}\Psi\supP{+}_n \\ \Psi\supP{-}_n \end{matrix}\Rs_\chi
 +c.c., 
\eequ
\end{widetext}
where $\chi$ and $\chi'(\neq\chi)$ denote the chiralities which we fix so that 
 $(\Psi\supP+)_\chi$ becomes even. 
Diagonalizing the above matrix and rearranging phases by the chiral rotation, 
 we find the KK spectrum is given as 
\bequ
 m_{KK}\supP{n}R=n+\theta^a (T_{\bf R_1}^a - T_{\bf R_2}^a)/\pi,\qquad n\in{\mathbb Z}.
\eequ
Let us note that the KK spectrum is basically the same form as the one 
obtained for the $S^1$ compactification, though the effect of $\theta^a$ is 
 non-trivial as discussed soon.
With this expression, it is easy to calculate the contributions to the one-loop 
 Higgs effective potential~\cite{Hosotani, S1}.

\begin{figure}[b]
\centering
\leavevmode
\includegraphics[width=8cm]{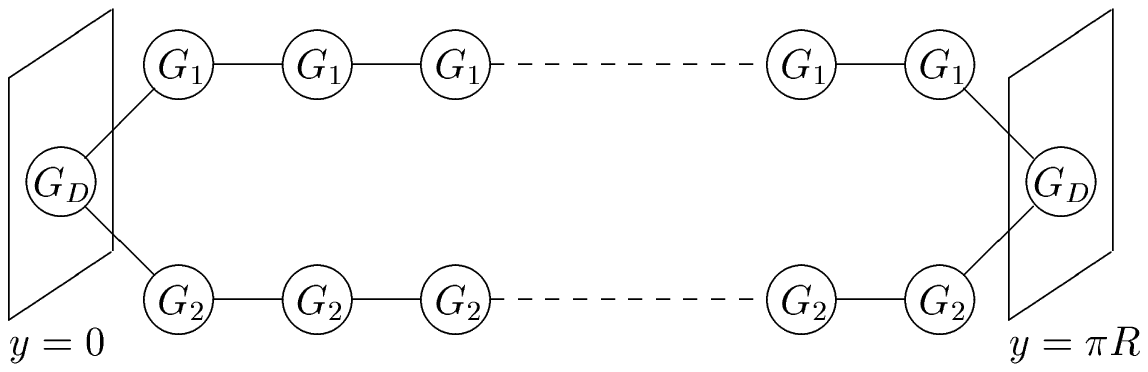}
\caption{The latticized picture of $G_1\times G_2\to G_D$ model on $S^1/\Z2$.}
\label{fig:deconstruction}
\end{figure}

It is interesting to see that for the fermions with ${\bf R_2}$ being trivial, 
the situation is completely the same as 
the $S^1$ compactification with the same radius $R$ and 
the original gauge coupling constant $g$. Note that the 4D effective 
gauge couplings are the same as that of the remaining $G_\diag$: 
$g_{\diag}^2/L_{S^1/\Z2}=g_{S^1}^2/L_{S^1}=g^2/(2\pi R)$.
This is understood by latticizing the extra-dimension~\cite{deconstruction} 
as shown in FIG.~\ref{fig:deconstruction}.
{}From this figure, we see the resulting Moose diagram looks the same 
 as the one for the $S^1$ model.
This means that what can be done in the $S^1$ models in the literatures 
are easily reproduced, while the chiral fermions can be put on the branes.

For the fermions with non-trivial ${\bf R_1}$ and ${\bf R_2}$, 
the mass spectra (and thus the contributions to the one-loop Higgs effective 
potential) are the same as the one of the fermion in the 
${\bf R_1}\times{\bf R_2}^*$ (reducible) representation in the $S^1$ model, 
while they behave as ${\bf R_1}\times{\bf R_2}$ representations under 
the remaining gauge symmetry $G_D$.
This gives a difference between our model and the $S^1$ model, which is 
again understood via the latticized picture 
as these fermions seem to have ``non-local" interactions 
with the gauge bosons in two distant sites in view of the $S^1$ side.

Special comments on $\psi$ are in order. As mentioned before, the 
above expressions are  
also for the case of $\psi$. Since the representation 
of $\psi_+$ is, however, different from that of $\psi_-$ in this case, it is 
clear that not all of the zero-modes acquire masses. 
Instead, some components that are vector-like with respect to the unbroken 
subgroup against the VEV of $A_5$ become massive. The other components 
are the eigenstates of $\theta^a(T_{\bf R_1}^a - T_{\bf R_2}^a)$ with 
vanishing eigenvalues and thus remain massless. These modes can be chiral.

Let us consider an example of an $SU(5)$ model with $\psi({\bf5})$ 
 where the VEV of $A_5$ breaks $SU(5)$ into $\GSM$. 
This $\psi$ is divided into $\psi_+({\bf15})$ and $\psi_-({\bf10})$, 
 which are decomposed as 
\beqn
 {\bf15}&\to&({\bf3},{\bf2})_{1/6}+({\bf6},{\bf1})_{-2/3}+({\bf1},{\bf3})_1,\\
 {\bf10}&\to&({\bf3},{\bf2})_{1/6}+({\bf{\bar3}},{\bf1})_{-2/3}+({\bf1},{\bf1})_1.
\eeqn
The component $({\bf3},{\bf2})_{1/6}$ becomes massive due to the VEV, 
 while the others remain massless. 
It is easy to see the latter are the eigenstates of 
 $\theta^a(T_{\bf R_1}^a - T_{\bf R_2}^a)$ with zero eigenvalues in this case.

\section{Applications}

In this section, we shall discuss applications to $SU(5)$ models. 
The literature~\cite{Hosotani,S1} indicates that it is not easy to realize 
vacua where the $SU(5)$ symmetry is broken to $SU(3)\times SU(2)\times U(1)$ 
as the global minimum of the potential in the $S^1$ framework. 
Since the potential in our setup is basically the same as those 
in the $S^1$ models, unfortunately, the same conclusion would 
be applied to our case. 

It is, however, not so important whether we reside on the global minimum or 
on a local minimum, as far as the lifetime of the local minimum is much longer 
than the age of universe. In this view, the analysis in 
Ref.~\cite{Sov} is noteworthy. There, it is claimed that the desired vacuum 
is realized as a local minimum when one fermion 
in ${\bf5}$ and ${\bf10}$, respectively, one scalar 
in $\bf5$ and three scalars in ${\bf15}$ are introduced.
Though we find that the precise point considered 
in Ref.~\cite{Sov} is not even extremal, there is another 
point, $\theta^aT^a/\pi={\rm diag}(2,2,2,-3,-3)/2$, being actually a 
local minimum where the desirable symmetry breaking occurs.
Since it is not easy to keep the scalars light against the quantum 
corrections, we would like to replace them by anti-periodic 
fermions which have a similar effect as the periodic scalars~\cite{APinGHU}. 
It is not difficult to check the point remains 
a local minimum after this modification.

In this way, we can utilize the results in the literature that 
investigate $S^1$ models while the chiral fermions can be put on 
the branes in our setup. We hope that this letter revives the 
researches in the literature that have been abandoned 
because of the lack of the chiral fermions. 
We leave concrete model building as future works%
\footnote{
In Ref.~\cite{gGHU-DTS}, for instance, it is claimed that the so-called 
 doublet-triplet splitting problem can be solved in supersymmetric version 
 of this scenario where light adjoint chiral multiplets are predicted. 
They have masses of the supersymmetry-breaking scale and thus are testable 
 at the TeV-scale collider experiments.
}.

\section*{Acknowledgments}

K.T. is 
supported in part by a Grant-in-Aid for Science Research (No. 21540285) from
the Japanese Ministry of Education, Science, Sports and Culture. 
The work of T.Y.\ was partially supported by the Japanese 
Society for the Promotion of Science.


\begin{thebibliography}{99}
%
\bibitem{Hosotani}
  Y.~Hosotani,
  Phys.\ Lett.\  B {\bf 126}, 309 (1983);
%
  Ann.\ of\ Phys.\  {\bf 190}, 233 (1989);
%
  Phys.\ Lett.\  B {\bf 129}, 193 (1983);
%
  Phys.\ Rev.\  D {\bf 29}, 731 (1984).
%
%
\bibitem{S1}
  A.~Higuchi and L.~Parker,
  Phys.\ Rev.\  D {\bf 37}, 2853 (1988);
%
  A.~T.~Davies and A.~McLachlan,
  Phys.\ Lett.\  B {\bf 200}, 305 (1988);
%
  Nucl.\ Phys.\  B {\bf 317}, 237 (1989);
%
  J.~E.~Hetrick and C.~L.~Ho,
  Phys.\ Rev.\  D {\bf 40}, 4085 (1989);
%
  A.~McLachlan,
  Nucl.\ Phys.\  B {\bf 338}, 188 (1990);
%
  C.~L.~Ho and Y.~Hosotani,
  Nucl.\ Phys.\  B {\bf 345}, 445 (1990);
%
  H.~Hatanaka,
  Prog.\ Theor.\ Phys.\  {\bf 102}, 407 (1999);
%
  K.~Takenaga,
  Phys.\ Lett.\  B {\bf 570}, 244 (2003);
%
  N.~Haba, K.~Takenaga and T.~Yamashita,
  Phys.\ Lett.\  B {\bf 605}, 355 (2005).
%
\bibitem{hierarchy}
N.~V.~Krasnikov,
Phys.\ Lett.\ B {\bf 273}, 246 (1991);
%
H.~Hatanaka, T.~Inami and C.~S.~Lim,
Mod.\ Phys.\ Lett.\ A {\bf 13}, 2601 (1998);
%
G.~R.~Dvali, S.~Randjbar-Daemi and R.~Tabbash,
Phys.\ Rev.\ D {\bf 65}, 064021 (2002);
%
N.~Arkani-Hamed, A.~G.~Cohen and H.~Georgi,
Phys.\ Lett.\ B {\bf 513}, 232 (2001);
%
I.~Antoniadis, K.~Benakli and M.~Quiros,
New J.\ Phys.\  {\bf 3}, 20 (2001);
%
%
 N.~Maru and T.~Yamashita,
 Nucl.\ Phys.\  B {\bf 754}, 127 (2006);
%
  Y.~Hosotani, N.~Maru, K.~Takenaga and T.~Yamashita,
  Prog.\ Theor.\ Phys.\  {\bf 118}, 1053 (2007).
%
%
\bibitem{S1Z2}
  C.~Csaki, C.~Grojean and H.~Murayama,
  Phys.\ Rev.\  D {\bf 67}, 085012 (2003);
%
  G.~Burdman and Y.~Nomura,
  Nucl.\ Phys.\  B {\bf 656}, 3 (2003);
%
  N.~Haba, M.~Harada, Y.~Hosotani and Y.~Kawamura,
  Nucl.\ Phys.\  B {\bf 657}, 169 (2003)
  [Erratum-ibid.\  B {\bf 669}, 381 (2003)];
%
  N.~Haba and Y.~Shimizu,
  Phys.\ Rev.\  D {\bf 67}, 095001 (2003)
  [Errata;\ {\bf 69}, 059902 (2004)];
%
  I.~Gogoladze, Y.~Mimura and S.~Nandi,
  Phys.\ Lett.\  B {\bf 560}, 204 (2003);
%
  C.~A.~Scrucca, M.~Serone and L.~Silvestrini,
  Nucl.\ Phys.\  B {\bf 669}, 128 (2003);
  R.~Contino, Y.~Nomura and A.~Pomarol,
  Nucl.\ Phys.\  B {\bf 671}, 148 (2003);
%
  K.~Agashe, R.~Contino and A.~Pomarol,
  Nucl.\ Phys.\  B {\bf 719}, 165 (2005);
%
 K.~Agashe and R.~Contino,
 Nucl.\ Phys.\  B {\bf 742}, 59 (2006);
%
  A.~D.~Medina, N.~R.~Shah and C.~E.~M.~Wagner,
  Phys.\ Rev.\  D {\bf 76}, 095010 (2007);
%
  Y.~Hosotani and Y.~Sakamura,
  Prog.\ Theor.\ Phys.\  {\bf 118}, 935 (2007);
%
  Y.~Hosotani, P.~Ko and M.~Tanaka,
  Phys.\ Lett.\  B {\bf 680}, 179 (2009);
%
  C.~S.~Lim and N.~Maru,
  Phys.\ Lett.\  B {\bf 653}, 320 (2007);
%
  N.~Haba, Y.~Sakamura and T.~Yamashita,
  JHEP {\bf 0907}, 020 (2009);
%
  JHEP {\bf 1003}, 069 (2010).
%
%
\bibitem{Kawamura}
 Y.~Kawamura,
 Prog.\ Theor.\ Phys.\  {\bf 103}, 613 (2000);
%
 ibid\  {\bf 105}, 691 (2001);
%
 ibid\  {\bf 105}, 999 (2001).
%
%
\bibitem{hetero}
  D.~J.~Gross, J.~A.~Harvey, E.~J.~Martinec and R.~Rohm,
  Phys.\ Rev.\ Lett.\  {\bf 54}, 502 (1985);
%
  Nucl.\ Phys.\  B {\bf 256}, 253 (1985);
%
  Nucl.\ Phys.\  B {\bf 267}, 75 (1986).
%
%
\bibitem{DiagonalEmbedding}
  K.~R.~Dienes and J.~March-Russell,
  Nucl.\ Phys.\  B {\bf 479}, 113 (1996).
%
%
\bibitem{heteroticGUTs}
  D.~C.~Lewellen,
  Nucl.\ Phys.\  B {\bf 337}, 61 (1990);
%
  G.~Aldazabal, A.~Font, L.~E.~Ibanez and A.~M.~Uranga,
  Nucl.\ Phys.\  B {\bf 452}, 3 (1995);
%
  J.~Erler,
  Nucl.\ Phys.\  B {\bf 475}, 597 (1996);
%
  Z.~Kakushadze and S.~H.~H.~Tye,
  Phys.\ Rev.\  D {\bf 55}, 7878 (1997);
%
  Phys.\ Rev.\  D {\bf 55}, 7896 (1997).
%
%
\bibitem{IKMMTTTY}
  M.~Ito {\it et al.},
  Phys.\ Rev.\  D {\bf 83}, 091703 (2011).
%
%
\bibitem{ExchangeBC}
 M.~Regis, M.~Serone and P.~Ullio,
 JHEP {\bf 0703}, 084 (2007);
%
 G.~Panico, E.~Ponton, J.~Santiago and M.~Serone,
 Phys.\ Rev.\  D {\bf 77}, 115012 (2008).
%
%
\bibitem{deconstruction}
  N.~Arkani-Hamed, A.~G.~Cohen and H.~Georgi,
  Phys.\ Rev.\ Lett.\  {\bf 86}, 4757 (2001).
%
%
\bibitem{Sov}
  V.~B.~Svetovoi and N.~G.~Khariton,
  Sov. J. Nucl. Phys. 43 (2), 280 (1986); 
  Yad.\ Fiz.\  {\bf 43} (1986) 438.
%
%
\bibitem{APinGHU}
  N.~Haba, Y.~Hosotani, Y.~Kawamura and T.~Yamashita,
  Phys.\ Rev.\  D {\bf 70}, 015010 (2004);
%
  N.~Haba, K.~Takenaga and T.~Yamashita,
  Phys.\ Lett.\  B {\bf 615}, 247 (2005);
%
  N.~Haba, S.~Matsumoto, N.~Okada and T.~Yamashita,
  JHEP {\bf 0602}, 073 (2006).
%
%
\bibitem{gGHU-DTS}
  T.~Yamashita,
  arXiv:1106.3229 [hep-ph].
%
%
\end{thebibliography}
\end{document}